\documentclass[12pt]{article}
\usepackage{epsfig}
\usepackage{epsfig}
\begin{document}
\def\tr{{\rm tr}\, }
\def\Tr{{\rm Tr}\, }
\def\hTr{\hat{\rm T}{\rm r}\, }
\def\be{\begin{eqnarray}}
\def\ee{\end{eqnarray}}
\def\ctt{\chi_{\tau\tau}}
\def\cta{\chi_{\tau a}}
\def\ctb{\chi_{\tau b}}
\def\cab{\chi_{ab}}
\def\cba{\chi_{ba}}
\def\ptt{\phi_{\tau\tau}}
\def\pta{\phi_{\tau a}}
\def\ptb{\phi_{\tau b}}
\def\>{\rangle}
\def\<{\langle}
\def\d{\hbox{d}}
\def\pab{\phi_{ab}}
\def\lb{\label}
\def\appendix{{\newpage\section*{Appendix}}\let\appendix\section%
        {\setcounter{section}{0}
        \gdef\thesection{\Alph{section}}}\section}
\renewcommand{\figurename}{Fig.}
\renewcommand\theequation{\thesection.\arabic{equation}}
\hfill{\tt }\\\mbox{} \vskip0.3truecm
\begin{center}
\vskip 2truecm {\Large\bf Entanglement entropy, conformal
invariance}\vskip 0.5truecm {\Large\bf and extrinsic geometry}
\vskip 1.5truecm {\large\bf Sergey N.~Solodukhin\footnote{ {\tt
solodukh@lmpt.univ-tours.fr}}
}\\
\vskip 0.6truecm \it{Laboratoire de Math\'ematiques et Physique
Th\'eorique CNRS-UMR 6083, \par Universit\'e de Tours, Parc de
Grandmont, 37200 Tours, France}
\end{center}
\vskip 1cm
\begin{abstract}
\noindent
We use the conformal invariance and the holographic
correspondence to fully specify the dependence of entanglement
entropy on the extrinsic geometry of the 2d surface $\Sigma$ that
separates two subsystems of quantum strongly coupled
${\mathcal{N}}=4$ $SU(N)$ superconformal gauge theory. We extend
this result and calculate entanglement entropy of a generic 4d
conformal field theory. As a byproduct, we obtain a closed-form
expression for the entanglement entropy in flat space-time when
$\Sigma$ is sphere $S_2$ and when $\Sigma$ is two-dimensional
cylinder. The contribution of the type A conformal anomaly to entanglement entropy
is always  determined by topology of surface $\Sigma$ while the dependence of the entropy
on the extrinsic geometry of $\Sigma$ is due to the type B conformal anomaly.

\end{abstract}
\vskip 1cm
\newpage

\section{Introduction}
\setcounter{equation}0

Entanglement entropy is a well-defined measure of the quantum
correlations between two subsystems separated by a surface
$\Sigma$. Since the correlations are short-range the entropy is
essentially determined by geometry, both intrinsic and as an
embedding into a larger space-time, of the surface $\Sigma$. The
recent studies revealed the important role the entanglement
entropy plays in the physics of black holes and in the quantum
field models \cite{Bombelli:1986rw}-\cite{Brustein:2005vx}. An
important element of the present understanding  is the holographic
interpretation of  entanglement entropy \cite{Ryu:2006bv},
\cite{Ryu:2006ef}. This interpretation suggests a purely geometric
way of computing the entropy of  a strongly coupled conformal
field theory \cite{Emparan:2006ni}-\cite{Hubeny:2007xt}.

In this note we give a derivation of  entanglement entropy of a
generic four-dimensional conformal field theory based on the
conformal properties of the theory. Our approach is partially motivated
by work of Dowker \cite{Dowker:1994bj}.
In particular, we suggest
that, in the large N limit,  the entanglement entropy calculated
in a four-dimensional ${\mathcal{N}}=4$ $SU(N)$ superconformal
gauge theory, the quantum field theory counterpart in the
holographic duality, is given by expression

\be S_{SC}={A(\Sigma)\over 4\pi\epsilon^2}N^2-{N^2\over
24\pi}\int_\Sigma (3R_{aa}-2R-{3\over 2} k_ak_a)\ \ln\epsilon
+s_{SC}(g)~~, \lb{1} \ee where $A(\Sigma)$ is area of $\Sigma$,
$R$ is the Ricci scalar of four-dimensional metric $g_{\mu \nu}$,
$R_{aa}=\sum_{i=1}^2 R_{\mu\nu}n^\mu_an^\nu_a$ is  projection of
the Ricci tensor onto the subspace orthogonal to  surface
$\Sigma$, $n^\mu_a,~ a=1,2$ is a pair of unite vectors orthogonal
to the surface $\Sigma$. $k_a$ is the trace of the second
fundamental form
$k^a_{\mu\nu}=-\gamma^\alpha_\mu\gamma^\beta_\nu\nabla_\alpha
n^a_\beta$, where $\gamma_{\mu\nu}=g_{\mu\nu}-n^a_\mu n^a_\nu$ is
the induced metric of $\Sigma$. $\epsilon$ is an UV cut-off needed
for regularization of the theory on curved background.

The logarithmic term in (\ref{1}) is  invariant under generic
conformal transformations, $g_{\mu\nu}\rightarrow e^{-2\omega}
g_{\mu\nu}$, in particular under those with non-vanishing normal
derivatives of $\omega$ on $\Sigma$. The term $k^a k^a$ in
(\ref{1}) takes care of the invariance  under these particular
transformations.

The function $s_{SC}(g)$ in (\ref{1}) is the UV finite part of the
entropy. Under global rescaling of the metric,
$g_{\mu\nu}\rightarrow \lambda^2 g_{\mu\nu}$, this function
changes as

\be s_{SC}(\lambda^2g)=s_{SC}(g)+{N^2\over 24\pi}\int_\Sigma
(3R_{aa}-2R-{3\over 2} k_ak_a)\ \ln \lambda ~~.\lb{2} \ee

Thus, the logarithmic term in (\ref{1}) and (\ref{2}) is due to
the integrated conformal anomaly. This aspect will be clarified
below. The property (\ref{2}) is particularly useful if the
configuration of surface $\Sigma$ in  4d space-time is
characterized by only one dimensionful parameter, $a$. Then,
rescaling the 4d metric and using  (\ref{2}) one can determine the
dependence of the entropy on this parameter. The logarithmic term
in entanglement entropy of a black hole was obtained in
\cite{Solodukhin:1994yz} and further studied in
\cite{Solodukhin:1994st}, \cite{Fursaev:1994te},
\cite{Mann:1997hm}, \cite{Solodukhin:1997yy}.

The expression (\ref{1}) is general in that it gives the entropy
for all possible choices of surface $\Sigma$. In particular,
$\Sigma$ can be a black hole horizon. The extrinsic curvature
$k^a, ~a=1,2$ vanishes in this case. The other possible choice of
$\Sigma$ is any compact surface in curved space-time. Of a
particular interest is a surface in flat space-time. Then, the
logarithmic contribution to the entropy is given entirely by the
extrinsic curvature of $\Sigma$.

The entropy (\ref{1}) has a holographic interpretation. In fact,
the holographic origin of the logarithmic term was already
verified in \cite{Ryu:2006ef} in the no black hole case and in
\cite{Solodukhin:2006xv} for a black hole horizon. In both cases
the extrinsic curvature of $\Sigma$ assumed to be
zero\footnote{Black hole horizon is a fixed point of abelian
isometry. The extrinsic curvature of horizon, thus, vanishes.}.
Thus,
 a goal of this note is to single out the important contribution
of the extrinsic curvature to the logarithmic term in (\ref{1}).
The holographic interpretation of entanglement entropy was
questioned recently in \cite{Schwimmer:2008yh}. Below we comment
on this.

That entanglement entropy may depend on the extrinsic curvature of
$\Sigma$ was earlier suggested in \cite{Mann:1996bi} based on the
heat kernel analysis of Dowker \cite{Dowker:1994bj}.

\section{Replica method, singular surface and the conformal structure of  effective action}
\setcounter{equation}0 The most efficient method to compute
entanglement entropy is to introduce at the surface $\Sigma$ a
conical singularity of small angle deficit $\delta=2\pi
(1-\alpha)$, compute the effective action $W(E^\alpha)$ on a
manifold $E^\alpha$ with a singular surface and then apply the
formula \be S=(\alpha\partial_\alpha-1)|_{\alpha=1}W(E^\alpha)
\lb{3.1} \ee and obtain the entanglement entropy. Explanation of
this method is available in papers \cite{Callan:1994py},
\cite{Calabrese:2004eu}.

In this note we are interested in entanglement entropy of a
four-dimensional conformal field theory. The effective action then
has a general structure \be W_{\rm CFT}(E^\alpha)={a_0\over
\epsilon^4}+{a_1\over \epsilon^2}+a_2\ln \epsilon
+w(g^{(\alpha)})~~. \lb{3.2} \ee The terms $a_0$ and $a_1$,
representing the polynomial UV divergences, are not  universal
while the term $a_2$ is universal and is determined by the
integrated conformal anomaly. $w(g^{(\alpha)})$ is the UV finite
part of the effective action. Under a global rescaling of the 4d
metric on $E^\alpha$, $g^{(\alpha)}\rightarrow \lambda^2
g^{(\alpha)}$, one has \be
w(\lambda^2g^{(\alpha)})=w(g^{(\alpha)})-a_2\ln \lambda~~.
\lb{3.3} \ee It is proved (see \cite{Dima} and references therein)
on a rather general grounds that for a generic manifold and a
generic conformal field  the corresponding coefficient $a_2$ is
conformal invariant, $a_2[e^{-2\omega}g]=a_2[g]$. This is an
important property that plays a key role in our consideration.

On 4d manifold with a singular surface $\Sigma$ the coefficients
$a_1$ and $a_2$ have both the bulk part and the surface part. To
first order in $(1-\alpha)$ one finds that \be
a_1(E^\alpha)=\alpha a^{\rm bulk}
_1(E)+(1-\alpha)a_1^\Sigma +O(1-\alpha)^2~~, \nonumber \\
a_2(E^\alpha)=\alpha a^{\rm bulk} _2(E)+(1-\alpha)a_2^\Sigma
+O(1-\alpha)^2~~, \lb{3.4} \ee where \be a_1^{\rm bulk}(E)=C\int_E
R~, ~~a_1^\Sigma=4\pi C\int_\Sigma 1 \lb{x} \ee
 with constant $C$
depending on the UV regularization scheme. In this note we use
normalization $C={1\over 16\pi^2}$.
 Coefficients $a_2^{\rm
bulk} (E)$ and $a_2(\Sigma)$ are respectively the integrated bulk
and surface conformal anomalies. Under a conformal transformation
$g\rightarrow e^{-2\omega}g$ one has that \be a_2^{\rm
bulk}(e^{-2\omega}g)=a_2^{\rm bulk}(g) \ {\rm and}  \
a_2^\Sigma(e^{-2\omega}g)=a_2^\Sigma (g)~~. \lb{3.5} \ee Applying
the formula (\ref{3.1}) one obtains entanglement entropy \be
&&S={a_1^\Sigma\over \epsilon^2}+a_2^\Sigma \ln \epsilon+s(g)~~,
\nonumber \\
&&s(\lambda^2 g)=s(g)-a_2^\Sigma\ln\lambda~~. \lb{3.6} \ee In four
dimensions the bulk conformal anomaly is a combination of two
terms, the topological Euler term and the square of the
Weyl tensor, \be &&a^{\rm bulk}_2=A E_{(4)}+B I_{(4)}~~,\nonumber \\
&&E_{(4)}={1\over 64}
\int_E(R_{\alpha\beta\mu\nu}R^{\alpha\beta\mu\nu}-4R_{\mu\nu}R^{\mu\nu}+R^2)~~,
\nonumber \\ &&I_{(4)}=-{1\over
64}\int_E(R_{\alpha\beta\mu\nu}R^{\alpha\beta\mu\nu}-2R_{\mu\nu}R^{\mu\nu}+{1\over
3} R^2)~~. \lb{3.7} \ee These are respectively the conformal
anomalies of type A and B. The surface contribution to the
conformal anomaly can be calculated directly by, for example, the
heat kernel method as in \cite{Fursaev:1994in}. The direct
computation although straightforward is technically involved. One
has however a short cut: there is a precise balance, observed in
\cite{Solodukhin:1994yz} and \cite{Fursaev:1994ea}, between the
bulk and surface anomalies, this balance is such that to first
order in $(1-\alpha)$ one can take $a_2(E^\alpha)=a_2^{\rm
bulk}(E^\alpha)+O(1-\alpha)^2$ and use for the Riemann tensor of
$E^\alpha$ the representation as a sum of regular and singular
(proportional to a delta-function concentrated on surface
$\Sigma$) parts. The precise expressions are given in
\cite{Fursaev:1994ea}, \cite{Fursaev:1995ef}. This representation,
however, is obtained under the assumption that the surface
$\Sigma$ has the extrinsic curvature vanishing. Under this
assumption one finds that \cite{Fursaev:1994ea},
\cite{Fursaev:1995ef}  \be &&a_2(E^\alpha)=\alpha a^{\rm bulk}
_2(E)+(1-\alpha)
a_2^\Sigma +O(1-\alpha)^2~~,\nonumber \\
&&a_2^\Sigma=A a_A^\Sigma+B a_B^\Sigma ~~,\nonumber \\
 &&a_A^\Sigma={\pi\over 8}\int_\Sigma
(R_{abab}-2R_{aa}+R) ~~,\nonumber \\
&&a_B^\Sigma= -{\pi\over 8}\int_\Sigma (R_{abab}-R_{aa}+{1\over
3}R)~~, \lb{3.8} \ee
where $R_{abab}=R_{\alpha\beta\mu\nu}n^\alpha_a n^\beta_b n^\mu_a n^\nu_b$,
$R_{aa}=R_{\alpha\beta}n^\alpha_a n^\beta_a$.

Each surface term in (\ref{3.8}) is invariant
under a sub-class of conformal transformations, $g\rightarrow
e^{-2\omega}g$, such that  the normal derivatives of $\omega$
vanish on surface $\Sigma$. The surface term due to the bulk
Euler number is, moreover, a topological invariant: using the
Gauss-Codazzi equation (\ref{2.5}) and in the assumption of
vanishing extrinsic curvature this term, as shown in
\cite{Fursaev:1995ef}, is proportional to the Euler number of the
2d surface $\Sigma$, \be a_A^\Sigma= {\pi\over 8} \int_\Sigma
R_\Sigma~~, \lb{3.9} \ee where $R_\Sigma$ is intrinsic curvature
of $\Sigma$.

As we discussed above the coefficient $a_2$ is conformal invariant
and thus it should be invariant under arbitrary conformal
transformations. The equation (\ref{3.8}) is however invariant
under a specific subclass of the conformal transformations, those
with the normal derivatives vanishing on $\Sigma$. This is because
the terms depending on the extrinsic geometry are neglected in
(\ref{3.8}). Such terms however become very important for the
invariance of $a_2$ on manifold with a ``squashed conical
singularity'' (in the terminology of Dowker \cite{Dowker:1994bj})
that has a structure of the product of a conical metric and the
surface $\Sigma$ scaled by an arbitrary conformal factor.

Thus, our goal in this section is to find a modification of
(\ref{3.8}) that would be invariant under a generic conformal
transformation with the normal derivatives of $\omega$
non-vanishing on $\Sigma$. A generalization can be easily found
with the help of formulae (\ref{2.4}), (\ref{2.5}) and (\ref{2.6})
presented in Appendix.

The terms with the normal derivatives of $\omega$ in the conformal
transformation of $a_2^\Sigma$ then can be cancelled by adding the
quadratic combinations of extrinsic curvature, $\tr k^2$ and $k_a
k_a$. We notice that the conformal covariant combination $(\tr
k^2-{1\over 2}k_ak_a)$ always can be added with a numerical
pre-factor, to be further specified. Additionally to the
requirement of the conformal invariance we demand that the surface
term in the anomaly determined by the Euler number to be
topological invariant (\ref{3.9}) even if the extrinsic curvature
of $\Sigma$ is non-vanishing. This latter condition completely
fixes the conformal covariant part in $a_A^\Sigma$. One then
obtains \be
&&a_2^\Sigma=A a_A^\Sigma+B a_B^\Sigma~~, \nonumber \\
 &&a_A^\Sigma={\pi\over 8}\int_\Sigma
(R_{abab}-2R_{aa}+R-\tr k^2+k_a k_a)={\pi\over 8}\int_\Sigma R_\Sigma~~, \nonumber \\
&&a_B^\Sigma= -{\pi\over 8}\int_\Sigma (R_{abab}-R_{aa}+{1\over
3}R +\mu (\tr k^2-{1\over 2} k_a k_a ))~~. \lb{3.10} \ee The
method used in this section to determine $a_2^\Sigma$ is that of
Dowker \cite{Dowker:1994bj}. The value of $\mu$ can not be
determined by the conformal invariance. In section 4 we use
the holographic correspondence in order to specify $\mu$\footnote{We assume
that $\mu$ does not depend on the field content of the theory.}.

The surface anomalies have been recently studied in
\cite{Schwimmer:2008yh} and a non-conformal invariant form of the type A
anomaly has been found. Although this issue requires a
further analysis we note that, as equations (\ref{3.7}) and
(\ref{3.10}) indicate, a simple form of the decomposition, like the one used in the cohomological
analysis of \cite{Schwimmer:2008yh}, of the Riemann tensor on
$E^\alpha$ on regular and singular parts is likely not valid when
the extrinsic curvature of $\Sigma$ is non-vanishing.

\section{Entanglement entropy of  ${\mathcal{N}}=4$ $SU(N)$ super Yang-Mills theory}
\setcounter{equation}0

Provided the value of constant $\mu$ in (\ref{3.10}) is determined
(this will be done in section 4) the result (\ref{3.10}) can be
used to calculate the entanglement entropy of any conformal field
theory in four dimensions.  In this section we consider a
particular conformal field theory that has a holographic dual
description in terms of the gravity on anti-de Sitter space-time.

In the large $N$ limit the ${\mathcal{N}}=4$ $SU(N)$
superconformal gauge theory is characterized by the conformal
anomaly, computed holographically in \cite{Henningson:1998gx},
that takes the form (\ref{3.7}) with \be A=B={N^2\over
\pi^2}~~.\lb{4.1} \ee On thus gets \be a_{2(SC)}^\Sigma={N^2\over
24\pi}\int_\Sigma (-3R_{aa}+2R+{3\over 2}k_ak_a- 3(\mu+1)(\tr
k^2-{1\over 2}k_ak_a))~~ \lb{4.2} \ee for the surface term. The
entanglement entropy of the superconformal gauge theory  then has
the form (\ref{3.6}) \be S_{SC}={A(\Sigma)\over
4\pi\epsilon^2}+a_{2(SC)}^\Sigma\ln\epsilon+s_{SC}(g)~~. \lb{33}
\ee

  As we show in the next section the holographic
correspondence predicts value $\mu=-1$. With this value the
entropy of the superconformal gauge theory is given by (\ref{1})
as announced in the introduction.

\section{The holographic calculation}
\setcounter{equation}0

In the holographic duality the quantum field theory is placed on a
regularized boundary (parametrized by $\epsilon$) of
five-dimensional anti-de Sitter space-time. The 2d  surface
$\Sigma$ that separates two subsystems in the entanglement entropy
calculation is, thus, defined on the regularized boundary. The
parameter $\epsilon$ plays the role of the UV cut-off on the
quantum field theory side. According to the proposal of Ryu and
Takayanagi \cite{Ryu:2006bv}, \cite{Ryu:2006ef}  in the anti-de
Sitter space-time one considers a minimal 3d surface $\Gamma$
which bounds the surface $\Sigma$. The quantity \be S={{\rm Area
}(\Gamma)\over 4G_N}~~, \lb{5.1}\ee where $G_N$ is
five-dimensional Newton's constant, then is equal to entanglement
entropy (\ref{33}) in the boundary quantum conformal field theory.
This proposal has been verified in many particular cases and there
are reasons why it should be valid in general
\cite{Fursaev:2006ih}. In this section we use this holographic
interpretation in order to determine the parameter $\mu$ in
(\ref{4.2}) and (\ref{3.10}).

To simplify things, we consider the case of flat four-dimensional
space-time. Then the logarithmic term in the entropy (\ref{4.2})
or (\ref{5.1})  is entirely due to the extrinsic curvature of
$\Sigma$.  We note that in flat space-time $R_\Sigma=k_a k_a-\tr k^2$,
as follows from (\ref{2.5}).

The result of the holographic calculation (\ref{5.1})
can be presented in general form \be S={A(\Sigma)\over
4\pi\epsilon^2}N^2-{N^2\over 24\pi}\int_\Sigma ({\gamma\over
2}k_ak_a+\beta \tr k^2)\ln\epsilon +..~~,\lb{55} \ee where ".."
stands for the finite part and we use that in the holographic
correspondence ${1\over G_N}={2N^2\over \pi}$.  We further
consider two choices of surface $\Sigma$, a two-dimensional
cylinder and sphere $S_2$, determine the constants $\gamma$ and
$\beta$ and then compare (\ref{55}) with (\ref{33}), (\ref{4.2})
to determine value of parameter $\mu$.

Generally, the holographic calculation of the logarithmic term in
(\ref{33}), (\ref{55}) is related to the surface anomalies studied
by Graham and Witten \cite{Graham:1999pm} (see also
\cite{Henningson:1999xi} and \cite{Berenstein:1998ij}).

\bigskip

\noindent{\it Two-dimensional cylinder}. We choose the AdS metric
in the form (and fix AdS radius $l=1$) \be ds^2={d\rho^2 \over
4\rho^2}+{1\over \rho}(-dt^2+dz^2+dr^2+r^2d\phi^2)~~. \lb{5.2} \ee
The surface $\Sigma$  corresponds to values of the coordinates
$\rho=0$, $r=a$, $0\leq \phi \leq 2\pi$, $0\leq z \leq L$, where
we introduced $L$ in order to regularize the length of the
cylinder. The surface $\Gamma$ is described by function $r(\rho)$.
It has area \be {\rm Area}(\Gamma)=2\pi
L\int_{\epsilon^2}{d\rho \over 2\rho^2}r(\rho)\sqrt{1+4\rho r'^2}~~,
\lb{5.3}\ee where $r'=\partial_\rho r(\rho)$.  The minimal surface
is described by equation \be \sqrt{1+4\rho
r'^2}=\rho^2\partial_\rho({4rr'\over \rho \sqrt{1+4\rho
r'^2}})~~.\lb{5.4}\ee For small $\rho$, the  asymptotic solution
of this equation  that approaches $\Sigma$ at $\rho=0$ is \be
r(\rho)=a-{\rho\over 4a}+O(\rho^2)~~.\lb{5.5} \ee Respectively,
the asymptotic behavior of the area is \be {\rm Area}(\Gamma)={\pi
L a}\int_{\epsilon^2}{d\rho \over \rho^2}(1-{1\over 8a^2}\rho
+..)=\pi L a ({1\over \epsilon^2}+{1\over 4a^2}\ln\epsilon+..)
~~.\lb{5.6} \ee Although expression (\ref{5.6}) is asymptotic the
finite part in (\ref{5.6}) can be easily restored.  Indeed, the
metric (\ref{5.2}) is invariant under rescaling, $\rho\rightarrow
\lambda^2\rho$, $t\rightarrow \lambda t$, $z\rightarrow \lambda z$
and $r\rightarrow \lambda r$. This invariance determines, up to an
additive numerical constant, the finite part. Thus, the
holographic entropy in the case of cylinder is \be
S={A(\Sigma)\over 4\pi\epsilon^2}N^2+{N^2\over 8}{L\over a}
(\ln{\epsilon\over a}+c_1)+c_2~~, \lb{5.7} \ee
where $c_1$ and $c_2$ are some constants.

In order to compare this to (\ref{55}) we have to know the
extrinsic curvature of the cylinder. In flat space-time one of the
normal vectors to surface $\Sigma$ is timelike, $n_1^t=1$. The corresponding
extrinsic curvature vanishes. The other normal vector is
$n_2^r=1$. It has the only non-vanishing component of the
extrinsic curvature $k^2_{\phi\phi}=a$. Thus, one has that \be
k_ak_a={1\over a^2}~,~~\tr k^2={1\over a^2}~~.\lb{5.8} \ee
Comparing now (\ref{5.7}) and (\ref{55}) we find a relation \be
\gamma+2\beta=-3~~.\lb{5.9}\ee

\bigskip

\noindent{\it Sphere $S_2$.} The AdS  metric is \be
ds^2={d\rho^2\over 4\rho^2}+{1\over
\rho}(-dt^2+dr^2+r^2(d\theta^2+\sin^2\theta d\phi))~~.\lb{5.10}
\ee The surface $\Sigma$ is defined by $\rho=0$, $r=a$, $0\leq
\phi\leq 2\pi$ and $0\leq \theta \leq \pi$. The surface $\Gamma$
is described by function $r(\rho)$ and has area \be {\rm
Area}(\Gamma)=4\pi \int_{\epsilon^2}{d\rho\over
2\rho^2}r^2(\rho)\sqrt{1+4\rho r'^2}~~.\lb{5.11}\ee The minimal
surface is described by equation \be r(\rho)\sqrt{1+4\rho
r'^2}=\rho^2\partial_\rho({2r^2r'\over \sqrt{1+4\rho
r'^2}})~~.\lb{5.12}\ee The  asymptotic solution takes the form \be
r(\rho)=a-{1\over 2a}\rho+O(\rho)^2~~. \lb{5.13} \ee The area of
the minimal surface is \be {\rm Area}(\Gamma)=2\pi
a^2\int_{\epsilon^2}{d\rho\over \rho^2}(1-{1\over 2a^2}\rho+..)=
{2\pi a^2\over \epsilon^2}+2\pi\ln \epsilon +.. ~~.\lb{5.14} \ee
The finite part in (\ref{5.14}) is determined by the invariance of
metric (\ref{5.10}) under rescaling, $\rho\rightarrow \lambda
\rho$, $t\rightarrow \lambda t$, $r\rightarrow \lambda r$.
Applying the holographic proposal one finds that entanglement
entropy in the case of sphere is (see also \cite{Ryu:2006ef})

\be S={A(\Sigma)\over 4\pi \epsilon^2}N^2+N^2\ln{\epsilon\over
a}+{c'}_2~~. \lb{5.15} \ee

Vector $n_1^t=1$ normal to $\Sigma$ has zero extrinsic curvature.
The other vector normal  to $\Sigma$, $n_2^r=1$, has the
non-vanishing extrinsic curvature $k^2_{\theta\theta}=a$,
$k^2_{\phi\phi}=a\sin^2\theta$. Hence, one has that \be
k_ak_a={4\over a^2}~,~~\tr k^2={2\over a^2}~~.\lb{5.16}\ee By
comparing (\ref{5.15}) with (\ref{55}) one finds that \be \gamma
+\beta=-3~~.\lb{5.17}\ee

\bigskip

Both relations, (\ref{5.9}) and (\ref{5.17}), are consistent if
\be \gamma=-3~,~~\beta=0~~.\lb{5.18}\ee The comparison of
(\ref{33}), (\ref{4.2})  with (\ref{55}) for values (\ref{5.18})
determines the value of the parameter $\mu$ in (\ref{4.2}) and
(\ref{3.10}), \be \mu=-1~~. \lb{5.19}\ee

\section{Generic 4d conformal field theory}
\setcounter{equation}0

The result (\ref{1}) for entanglement entropy is easily
generalized for  a generic conformal field theory in four
dimensions. The conformal anomaly of such a theory is a
combination of the type A and type B anomalies, the surface term
taking the form (\ref{3.10}) with value of parameter $\mu=-1$,
already determined in section 4 by using the holographic correspondence.
Entanglement entropy of the theory is then given by (\ref{3.6}),
\be S_{(A,B)}= {A(\Sigma)\over 4\pi
\epsilon^2}+(Aa_A^\Sigma+Ba_B^\Sigma)\ln\epsilon +s_{(A,B)}(g)~~,
\lb{6.1}\ee where  $a_A^\Sigma$ and $a_B^\Sigma$ are given by
(\ref{3.10}) (with $\mu=-1$). The contribution of the type A anomaly is always
determined by topology of surface $\Sigma$ while the dependence of the entropy
on the extrinsic geometry of $\Sigma$ is due to the type B anomaly.

In flat space-time only the extrinsic curvature contributes to
$a_2^\Sigma$ and one has that \be S_{(A,B)}= {A(\Sigma)\over
4\pi\epsilon^2}+{\pi\over 8}\left(A\int_\Sigma
R_\Sigma+B\int_\Sigma (\tr k^2-{1\over 2}k_ak_a)
\right)\ln\epsilon +s_{(A,B)}(g)~~,\lb{6.2}\ee
where $R_\Sigma=k_ak_a-\tr k^2$ in flat space-time.
The finite part $s_{(A,B)}$ in
(\ref{6.2}) can be determined, using the rescaling property of
$s_{(A,B)}$, for some  geometries, in particular for sphere $S_2$
and the two-dimensional cylinder. Using the expressions, obtained
in section 4, for the extrinsic curvature of these two surfaces in
flat space-time one finds that the entanglement entropy of a
generic CFT in the case of cylinder is \be S_{(A,B)}^{\rm
cylinder}= {A(\Sigma)\over 4\pi\epsilon^2}+B{\pi^2\over 8} {L\over
a}\ln{\epsilon\over a} \lb{6.3}\ee and \be S_{(A,B)}^{\rm
sphere}={A(\Sigma)\over 4\pi \epsilon^2}+A{\pi^2}
\ln{\epsilon\over a} ~~\lb{6.4}\ee in the case of sphere $S_2$.
Thus, these two geometries single out two different types of conformal
anomaly in the calculation of entanglement entropy.

\section{Conclusions}
\setcounter{equation}0 The dependence of entanglement entropy on
the extrinsic geometry of the surface $\Sigma$ that separates two
subsystems is a missing element in the previous study of
entanglement entropy (see, however, \cite{Mann:1996bi} for an
earlier discussion). In particular, the extrinsic geometry plays
an important role in the entanglement entropy calculation in flat
space-time. In this paper we use the conformal symmetry and the
holographic correspondence and fully specify the way the extrinsic
curvature appears in  entanglement entropy. The case of the
conformal field theory dual to a gravity on anti-de Sitter
space-time is considered in detail. We extend this consideration
to a generic conformal field theory in four dimensions. In
particular, we obtain a closed-form expression for entanglement
entropy of a generic CFT in the case when $\Sigma$ is a
two-dimensional cylinder and when $\Sigma$ is two-dimensional
sphere. We specify the way the type A and the type B conformal anomalies
show up in the entanglement entropy.

\bigskip

\bigskip

\noindent {\large \bf Acknowledgments} \\
\vskip 2mm
\noindent I thank D. Fursaev for
helpful remarks and S. Theisen for email communication.

\bigskip

\noindent

\appendix{Conformal symmetry and the extrinsic geometry}
\setcounter{equation}0

In this section we collect the useful formulas for the conformal
transformation of various geometric quantities defined on a
 co-dimension 2 submanifold $\Sigma$.

 Under the transformation $g_{\mu\nu}\rightarrow e^{-2\omega}
 g_{\mu\nu}$, $n_\mu^a\rightarrow e^{-\omega}n^a_\mu$, $a=1,2$ the extrinsic
 curvature $k^a_{\mu\nu}=-\gamma^\alpha_\mu\gamma^\beta_\nu\nabla_\alpha
n^a_\beta$, $k^a=\gamma^{\mu\nu}k^a_{\mu\nu}$ of $\Sigma$
 changes as follows

\be k^a_{\mu\nu}\rightarrow
e^{-\omega}(k^a_{\mu\nu}+\gamma_{\mu\nu} n_a^\alpha \nabla_\alpha
\omega)~~, \nonumber \\
k_a\rightarrow e^{-\omega}(k_a+(d-2) n^\alpha_a\nabla_\alpha
\omega)~~. \lb{2.1} \ee

The projections of the Riemann and Ricci tensors,
$R_{abab}=R_{\alpha\beta\mu\nu}n^\alpha_a n^\beta_b n^\mu_a n^\nu_b$,
$R_{aa}=R_{\alpha\beta}n^\alpha_a n^\beta_a$, on the subspace
orthogonal $\Sigma$ transform as \be &&R_{abab}\rightarrow
e^{2\omega}(R_{abab}+2n_a^\alpha n_b^\beta
\nabla_\alpha\nabla_\beta \omega +2(n_a^\alpha\partial_\alpha
\omega)(n_a^\alpha\partial_\alpha \omega)-2(\nabla \omega)^2)~~,
\lb{2.2} \\
&&R_{aa}\rightarrow e^{2\omega}(R_{aa}+(d-2)n_a^\alpha n_a^\beta
(\nabla_\alpha \nabla_\beta \omega +\nabla_\alpha \omega
\nabla_\beta \omega ) +2(2-d) (\nabla \omega )^2+2\Delta \omega
)~~. \nonumber
%&&R\rightarrow e^{2\omega}(R+{(d-1)\over 2}(4\Delta \omega
%+2(2-d)(\nabla \omega)^2)
  \ee
The Laplace operator $\Delta$ is further presented in the form \be
\Delta \omega=\hat{\Delta}\omega +n_a^\alpha n_a^\beta
\nabla_\alpha\nabla_\beta \omega -k_a n_a^\alpha \partial_\alpha
\omega ~~,\lb{2.3} \ee where $\hat{\Delta}$ is the intrinsic
Laplacian on $\Sigma$.

In dimension $d=4$ there are two conformal covariant combinations of
the Riemann curvature and the extrinsic curvature
\cite{Dowker:1994bj},
 \be &&(2R_{abab}-R_{aa}-{1\over 2} k_a k_a) \rightarrow e^{2\omega}(2R_{abab}-R_{aa}-{1\over 2}k_ak_a-2\hat{\Delta}\omega)~~,
 \nonumber \\
&& (\tr k^2-{1\over 2}k_ak_a)\rightarrow e^{2\omega}(\tr
k^2-{1\over 2}k_ak_a) ~~.\lb{2.4} \ee Using the Gauss-Codazzi
equation \be R=R_\Sigma +2R_{aa}-R_{abab}-k_ak_a+\tr
k^2~~,\lb{2.5} \ee where $R_\Sigma$ is the intrinsic Ricci scalar
of the surface, one finds that the conformal covariant combination
involving the Ricci scalar is \be (3R_{aa}-2R-{3\over
2}k_ak_a)=(2R_{abab}-R_{aa}-{1\over 2}k_ak_a)-2R_\Sigma +2({1\over
2}k_ak_a-\tr k^2)~~. \lb{2.6} \ee

\end{document}